# Battery testing methods in fuel cell research

Shahbaz Ahmad[1,2] and Mehmet Egilmez[1,2]

[1] Department of Physics, American University of Sharjah, Sharjah

United Arab Emirates, PO Box: 26666

[2] Material Science and Engineering Ph.D. Program, American University of Sharjah, Sharjah,

United Arab Emirates, PO Box: 26666

## Abstract

This report presents some of the key laboratory electrochemical battery testing methods that are used in fuel cell research. Methods such as voltammetry, chronoamperometry, chronopotentiometry, and electrochemical impedance spectroscopy are of major importance. All the electrochemical corrosion tests are performed through a tri-electrode polarization cell setup containing a reference electrode, a counter electrode, and the working electrode (metal sample of interest) in an electrolyte solution. All three electrodes are connected to a potentiostat. Corrosion of metal occurs through an oxidation-reduction (redox) reaction. All the above testing methods can be performed by manipulating the current and voltage responses from the cell. Potentiostatic experiments (voltammetry and chronoamperometry) are performed using constant potential at the working electrode and recording the current response while galvanostatic experiments (chronopotentiometry) and vice versa. The measured data through all these experiments can provide very useful information regarding reaction reversibility, diffusion coefficient, reduction potential, rate of chemical reaction, durability, adsorption, voltage losses, and effective resistance to the mass and charge transport offered by electrode material.

## 1. Introduction.

Batteries and fuel cells are constructed on the common idea of basic galvanic cells or voltaic cells.

### 1.1. Galvanic Cells.

Galvanic cells are electrochemical cells that convert chemical energy into electrical energy through spontaneous reduction and oxidation (redox) reactions [1]. A typical redox reaction consists of transferring electrons resulting from the oxidation of substance on the negative electrode (anode) through the external circuit to the positive side (cathode) where it completes the reduction process by combining with another substance. The potential that occurred during the reaction which is also known as standard potential ($E°$) is determined by finding the difference between the cathodic ($E°_{Cathode}$) and anodic ($E°_{anode}$) potential resulting from two half-reactions [2]. However, we cannot find the potential of an electrode directly. We must use a reference electrode whose potential is at 0 V under standard conditions. The standard hydrogen electrode (SHE) is used for that purpose which has been assigned a potential value of 0 V under standard conditions (temperature of 298 K, 1M acidic solution, and 1 atm pressure) [3]. SHE consists of a wire strip made of highly conductive metals such as gold (Au) or platinum (Pt) connected to the platinized surface inside an aqueous solution containing 1 M $H^+$ ions. Hydrogen ($H_2$) gas is supplied to the solution at 1 atm which maintains a constant equilibrium with $H^+$ ions on the Pt-solution interface. Under these standard conditions, the $H^+$ ions are reduced in $H_2$ molecules on the Pt surface according to the following reaction.

$$2H^+_{(aq)} + 2e^- \rightleftharpoons H_{2(gas)} \tag{1}$$

Figure 1 shows a galvanic cell composed of a Zinc (Zn) anode immersed in an aqueous solution of zinc nitrate, $Zn(No_3)_2$, and SHE cathode in another beaker. Both beakers are connected through a circuit connected with a voltammeter for measuring potential differences across both electrodes. In a closed circuit, the zinc anode starts to dissolve in $Zn^+$ ions while on the cathode side hydrogen begins to reduce in $H_2$. During this redox reaction, the voltammeter indicates a potential of 0.76 V. The corresponding half-reactions that occurred at both electrodes are as follows.

Anode

$$Zn_{(s)} \longrightarrow Zn^{2+}_{(aq)} + 2e^- \qquad E°_{anode} = -0.76 \text{ V} \tag{2}$$

Cathode

$$2H^+_{(aq)} + 2e^- \longrightarrow H_{2(g)} \qquad E°_{Cathode} = 0.00 \text{ V} \tag{3}$$

Overall

$$Zn_{(s)} + 2H^+_{(aq)} \longrightarrow Zn^{2+}_{(aq)} + H_{2(g)} \tag{4}$$

$$E° = E°_{Cathode} - E°_{anode} = 0 - (-0.76) = 0.76 \text{ V} \tag{5}$$

So the standard potential that occurred at the anode during oxidation of Zn to $Zn^{2+}$ is called conventionally known as reduction cell potential ($E°_{Cell}$) is 0.76 V. Further the cell potential under nonstandard conditions ($E$) is determined from $E°$ through the famous Nernst equation as follows.

$$E = E° - \frac{0.0592 \text{ V}}{n} \log_{10} Q \tag{6}$$

Where $E$ = potential at nonstandard conditions (V), $E°$ = potential at standard conditions (V), $n$ = number of electrons transferred during reaction, $Q$ = reaction quotient (determines the direction of reaction)

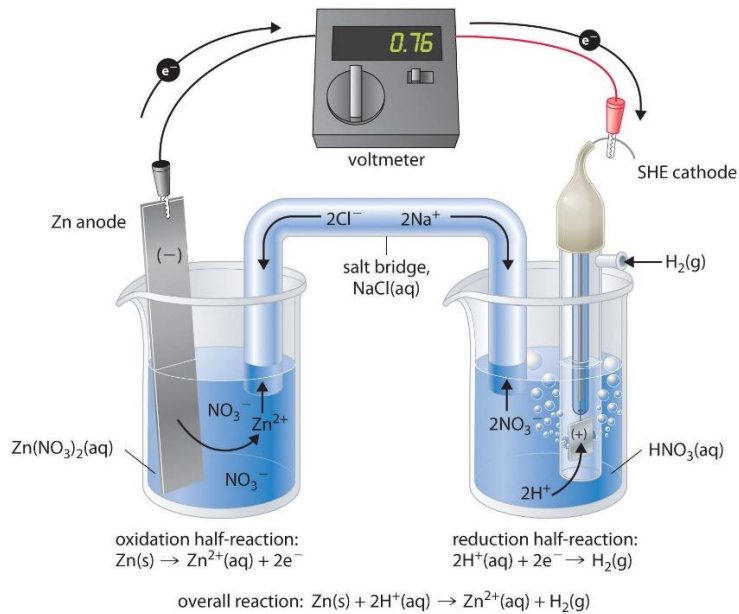

**Figure 1.** *Calculating standard electrode potential of Zinc electrode using SHE* [4]

### 1.2. Batteries

Batteries are engineered galvanic cells that use highly concentrated solid paste as an electrolyte instead of a solution to achieve maximum electric output per unit mass. The reason for this is to maintain the concentration of both reactants and products during the discharging/charging process thus resulting in constant output voltage. A series of galvanic cells are connected in such a fashion that the positive terminal of each cell is connected to the negative terminal of the other so the output voltage is the arithmetic sum of all individual cell voltages. Batteries are classified into two main types: primary batteries or irreversible batteries (figure 2a and b) which are disposable means that the electrode reactions are completely irreversible and secondary or reversible batteries (figure 2c) which are rechargeable and the insoluble product that adheres to the electrode during discharging process can be separated by recharging it. During the recharging process, batteries act as an electrolytic cell. An electrolytic cell uses an external electric energy source to decompose a substance through a reverse redox reaction called electrolysis. The direction of the reaction is reversed by changing the polarity of electrodes. Figure 3 shows the basic differences between galvanic and electrolytic cells.

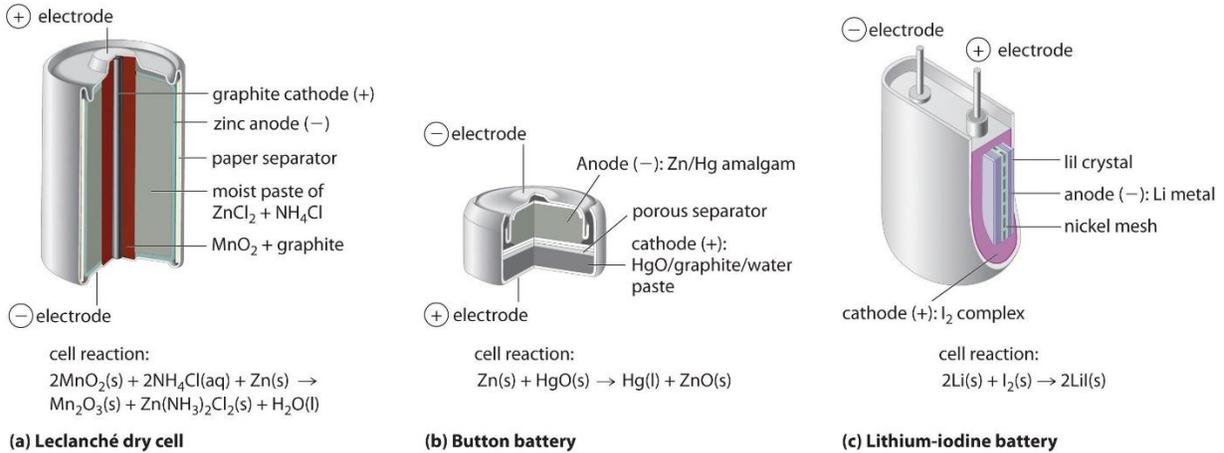

**Figure 2.** *Battery types (a, b) Irreversible type or Primary batteries, (c) Reversible type or secondary battery* [5]

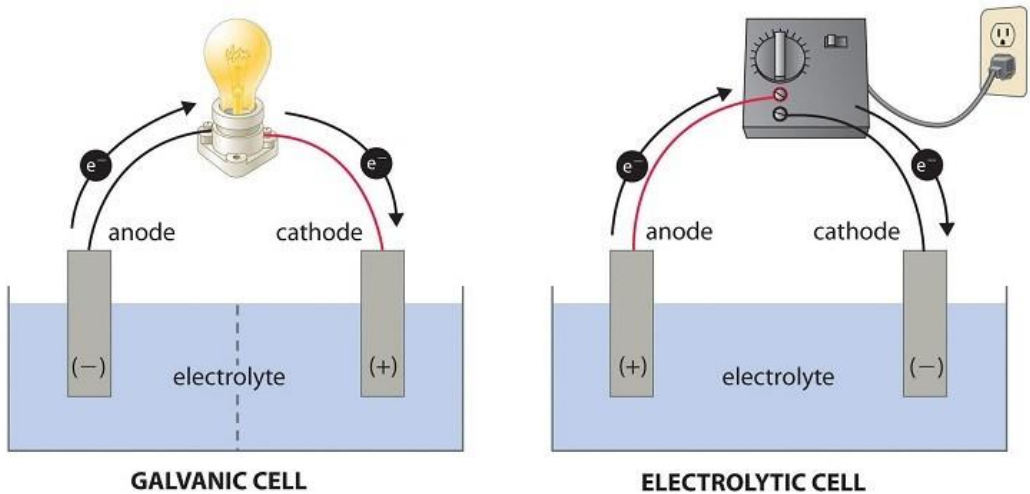

**Figure 3.** *Galvanic cell (left) and electrolytic cell (right)* [6]

### 1.3. Fuel Cells

Fuel Cells are another type of galvanic cell that uses hydrogen gas as fuel to produce electric energy. A constant supply of gas reactants is needed to convert the chemical energy from $H_2$ to electric energy through spontaneous redox reactions. Hydrogen is supplied at the anode (-) side which is oxidized thus creating free electron and proton [$H^+$]. The free electron runs through the external circuit while the $H^+$ passes through the semipermeable polymeric membrane to the cathode side (+) where it combines with the Oxygen and the incoming free electron to make water ($H_2O$). The net reaction is exothermic and the corresponding reactions at both anode and cathode are presented as under.

Anode

$$H_{2\,(g)} \longrightarrow 2H^+ + 2e^- \qquad E_o = 0.00 \text{ V} \qquad (7)$$

Cathode

$$O_{2(g)} + 4e^- + 4H^+ \longrightarrow 2H_2O_{(g)} \quad E_o = 1.23 \text{ V} \quad (8)$$

Net reaction

$$H_2 + 1/2 O_2 \longrightarrow H_2O_{(g)} \quad \Delta G_f = -229 \text{ kJ/mol} \quad (10)$$

Unlike a battery, it does not store chemical or electric energy as the energy from the incoming fuel is extracted directly through chemical reactions. Hydrogen fuel cells use acidic or basic electrolytes as ion carriers. The electrolytes are solid polymers with selection adsorption such as proton exchange membranes (acidic fuel cells) or anion exchange membranes (alkaline fuel cells) etc. Commercial hydrogen fuel cells use highly dispersed Pt on large surface area carbon (commercial Pt/C) as an electrode for hydrogen oxidation and oxygen reduction reactions (HOR and ORR). Fuel cells have been used in many applications such as US manned space vehicles, automobiles, power plants, ships, etc. The major hindrance that keeps this technology away from commercial use is the high cost of corroding noble metals that are used as electrode catalysts to boost sluggish oxygen reduction reactions. According to the US department of energy, the cost of commercially used Pt is almost 48% of full-stack commercial fuel cells [7]. Figure 4 shows a detailed schematic of a hydrogen fuel cell.

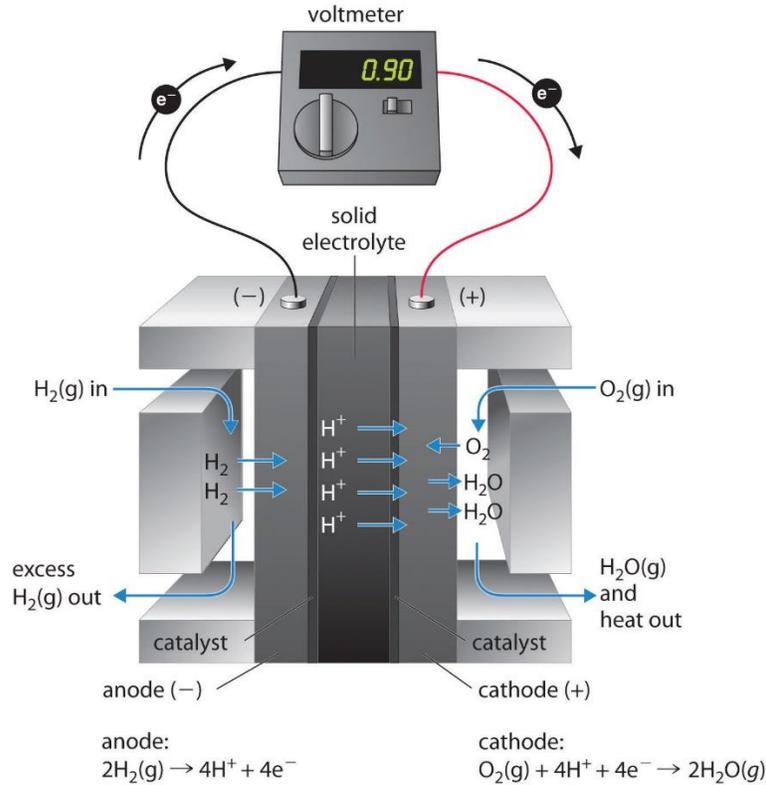

**Figure 4.** *Hydrogen fuel cell* [8]

## 2. Experimental Methods and Discussion
## 2.1. Electrochemical testing methods in batteries and fuel cells

According to the Nernst equation, the effective power produces by all types of galvanic cells are in relation to the different parameters such as current density, the concentration of electroactive species, activation energy or activation potential, number of valance electrons transfer (n), ions or mass transport in an electrolyte, internal resistance offered by electric connections, electrolyte, tolerance to residue or product formation, catalysts durability, etc. To address these terms several laboratories, and electrochemical testing methods are performed which are common among fuel cell batteries. Almost all of these tests are performed through a rotating disk electrode (RDE) system connected with a potentiostat.

Rotating disk electrode (RDE) is a hydrodynamic working electrode that uses a tri-electrode system to investigate the redox reaction mechanism of the catalyst materials [9]. An important idea behind the RDE system is the method through which an electroactive species move to the surface of the working electrode through convective diffusion where they are either oxidized or reduced [10]. The three-electrode system consists of a working electrode which consists of conductive metal embedded in insulating material as shown in figure 5a. The conductive materials are made up of glassy carbon, Pt, or Au. The working electrode is assembled with the mental shaft which rotates with variable rpm through an electric motor. To prepare a working electrode for electrochemical analysis, a catalyst material is mixed into Isopropyl alcohol along with Nafion dispersion as a binder and dipped onto the glassy carbon (GC) surface. The material is allowed to adhere to the GC surface for 20 minutes by rotating the working electrode at 400 rpm under ambient temperature. Finally, the dried working electrode which acts as a cathode (+) is dipped into a beaker containing Pt wire as a counter electrode (-) and a silver/silver chloride (Ag/AgCl) strip of saturated hydrogen electrode (SHE) discussed above as reference electrode [11]. During the working condition, the charge transfer occurs between a working electrode and counter electrode while the potential is measured with reference to a reference electrode. All three electrodes are connected to a potentiostat for the electroanalytical analysis of data. The beaker is filled with electrolytes (acidic or basic) along with two holes for the supply of reactant gases ($H_2/N_2$ and $O_2$). Figure 5b shows a complete setup for RDE analysis.

The use of RDE in electrochemical science is very wide but here we will discuss some of the common testing methods that are used for both batteries and fuel cell electrodes. The following testing methods are.

## 2.2. Voltammetric Methods
Voltammetry is an electroanalytical technique through which information about the analyte is obtained based on the current time response of a working electrode with respect to the applied potential [12]. The curves $I = f(E)$ obtained from data are called voltammograms. In a typical RDE instrument, a variable voltage profile from the potentiostat is applied to the working electrode with respect to the reference electrode (Ag/AgCl or SHE), and the total current generated as a result of the electrode reaction is measured. The profile of voltammograms depends upon the variation of potential per unit time (V/t) called scan rate and on the mass transfer during chemical reaction

[13]. The potential applied to the working electrode can be oxidizing or reducing depending on the type of analysis you are doing. The positive potential applied to the working electrode indicates the oxidation process while negative potential shows reduction.

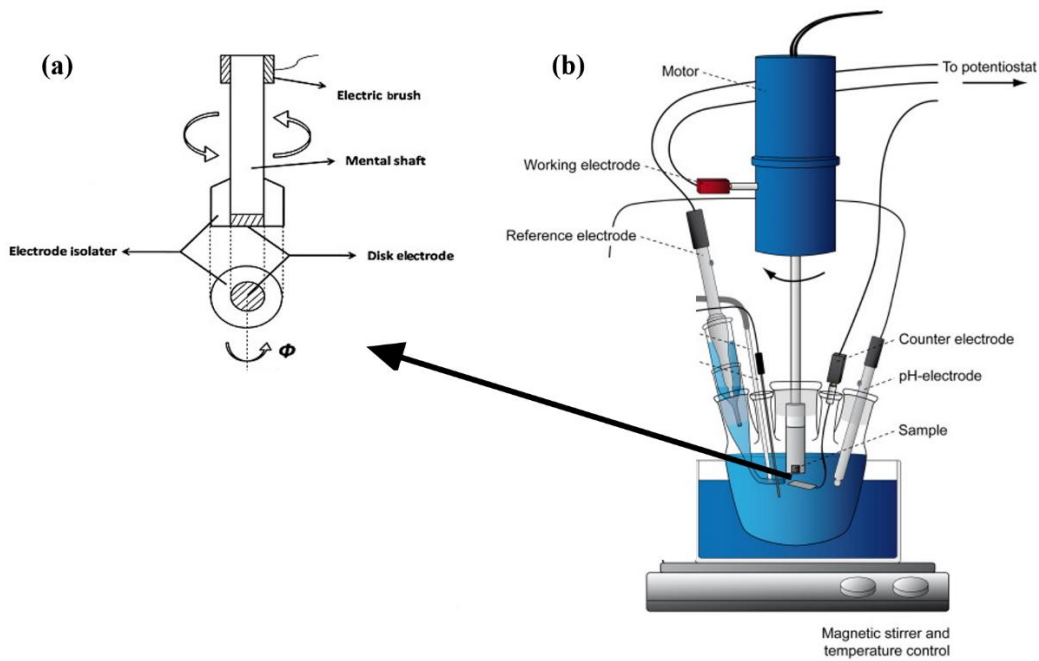

**Figure 5.** *Rotating disk electrode (a) Working electrode, (b) Complete setup* [9,11]

The two most common voltammetric methods that are common in both batteries and fuel cells are cyclic voltammetry, and linear sweep voltammetry.

### 2.2.1. Cyclic Voltammetry (CV)

Cyclic voltammetry (CV) is a primary electroanalytical method for determining the kinetics of redox reactions in electrochemical systems. CV is performed by cycling the potential of the working electrode and measuring the current obtained using a wide range of scan rates i.e. 10 – 10,000 V/s. The potential $E_1$ applied to the working electrode is measured against a reference electrode $E_2$ e.g. SHE which maintains a constant value of 0V. The applied working potential $E_1$ causes an excitation signal which can be in a forward or reverse direction. In the forward scan, the potential scans in a negative direction (reduction process) starting from negative maximum (a) in figure 6a and b and ends at lower extrema (d). Point (d) is known as the switching potential or reversing point where the analyte has completed the oxidation or reduction process. From point (d) to (g) the potential scans in a positive direction (oxidation process). The process of oxidation and reduction depends on the type of analyte used in the electrochemical system. Some analyte undergoes oxidation process first followed by reduction and vice versa. The slope of the excitation line determines the scan rate used in the process. Consider a reversible redox reaction of a single electron.

$$A^+ + e^- \rightleftharpoons A \tag{11}$$

CV profile of working electrode is obtained by plotting the measured current against the potential scan. Figure 6b shows cyclic voltammograms for equation (11).

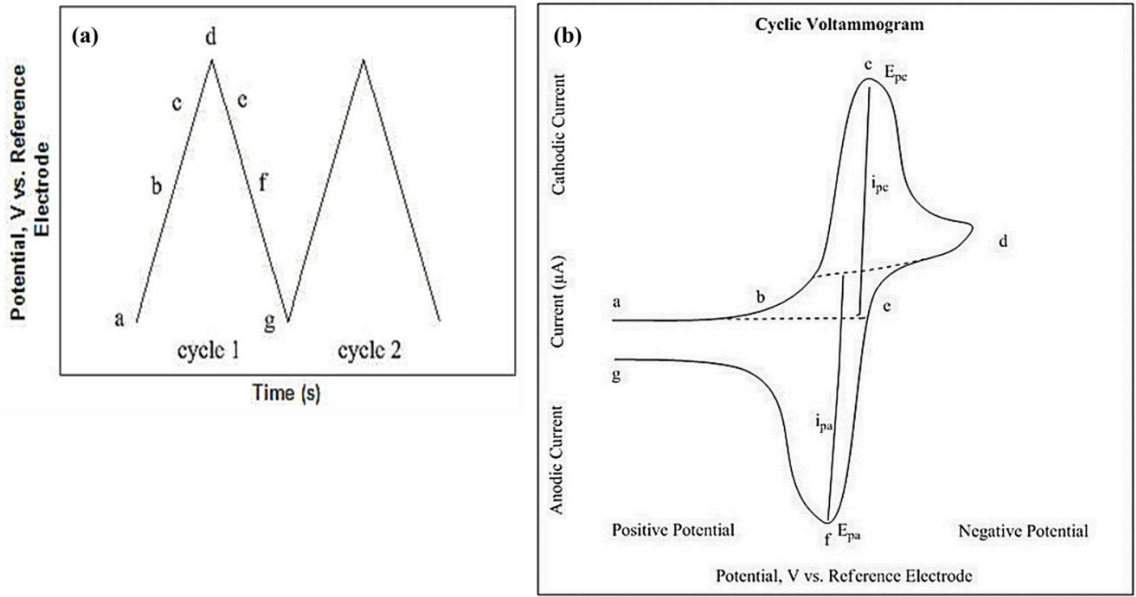

**Figure 6.** *Cyclic Voltammetry profiles (a) Excitation signal of scan (b) Redox voltammograms of single-electron* [14]

A negative potential scan causes a reduction current process starting from negative maxima (a) to switching potential (d). In this region the working potential scans negatively causing a reduction process therefore the current generated in this region is called cathodic current ($I_c$) and the current measured at point (c) due to cathodic peak potential ($E_{pc}$) is called cathodic peak current ($I_{pc}$) which is given by equation 12 [1].

$$I_{pc} = 0.4463nFCA(Da)^{1/2} \quad (12)$$

Where $n$ = number of electrons transfer in a reduction, F = Faraday constant (Q/mol), C = molar concentration of electroactive species (mol/dm³), D = Diffusion coefficient of electroactive species (cm²/s), A = surface area of the working electrode (cm²) and a is given by equation 13

$$a = nFv/RT \quad (13)$$

Where $v$ = sweep rate (V/s), $R$= Universal gas constant (J/mol.k), $T$ = Temperature (K).

By the increasing potential of the working electrode in the negative direction, the reduction process increases till $E_{pc}$ at point (c) where all the electroactive species have been reduced. Further increasing potential beyond $E_{pc}$ does not affect current because the concentration of product accumulation in electrode area increases which block the diffusion of species toward electrode (figure 7) and hence cathodic current starts to decrease till (d) which is called switching or reversing point. From this point onwards the potential scans in a positive direction and oxidation reaction occur till (f) where all the substrate have been oxidized. The resulting current from the oxidation process is called an anodic current and the current obtained at (f) corresponding to anodic peak potential $E_{pa}$ is called anodic peak current $I_{pa}$.

The potential applied to the working electrode during oxidation or reduction is given by.

$$E = E_i + vt \qquad (14)$$

$E_i$ = Initial potential (V), t = time (sec)

When the direction of applied working potential is reversed.

$$E = E_s - vt \qquad (15)$$

$E_s$ = Potential at the switching point

The reduction potential ($E_o$) measured during CV is the mean of potential at anodic and cathodic peaks ($E_{pa}$ and $E_{pc}$).

$$E_o = (E_{pa} + E_{pc}) / 2 \qquad (16)$$

CV can be used for many applications in electrochemical systems such as finding intermediates in a redox reaction, electron transfer pathway (some materials prefer 2e$^-$ pathway while other 4e$^-$ to make $H_2O$ product in hydrogen fuel cell depending upon the electron stoichiometry in catalyst material), reaction reversibility, reduction potential, diffusion coefficient and concentration of an unknown analyte through a reversible Nernst system in which concentration of a solution is directly proportional to the reduction current.

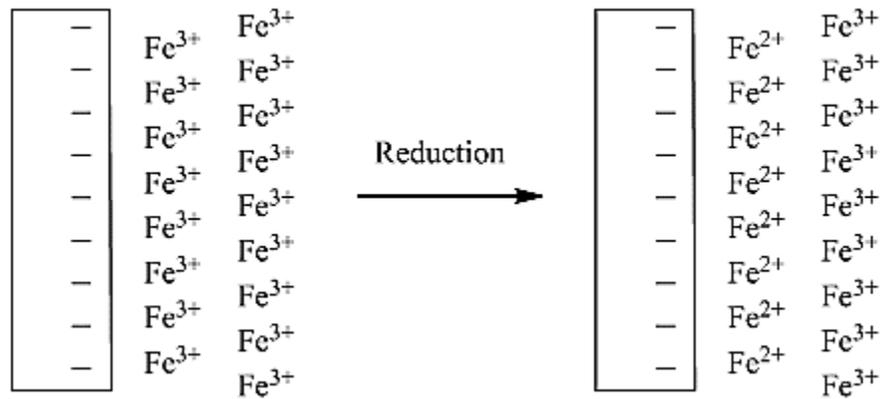

**Figure 7.** *Complete cationic reduction of $Fe^{3+}$ to $Fe^{2+}$ near a working electrode in cyclic voltammetry* [15]

### 2.2.2. Linear Sweep Voltammetry (LSV)

Linear sweep voltammetry (LSV) is a potentiostatic subset linear sweep method of basic cyclic voltammetry. Unlike CV the potential applied to working electrode $E$ in LSV is scanned linearly between two points $E_i$ and $E_f$ (figure 8a) and the current generated is measured as a function of time [16]. LSV is preferred over CV in a case when the reaction is irreversible and we are not able to extract useful information from the return cycle. In other words, the entropy generated during a chemical reaction is large enough to make the process irreversible [17]. LSV is a unidirectional and rapid technique as compared to a CV which provides both qualitative and quantitative analysis of an electrochemical system.

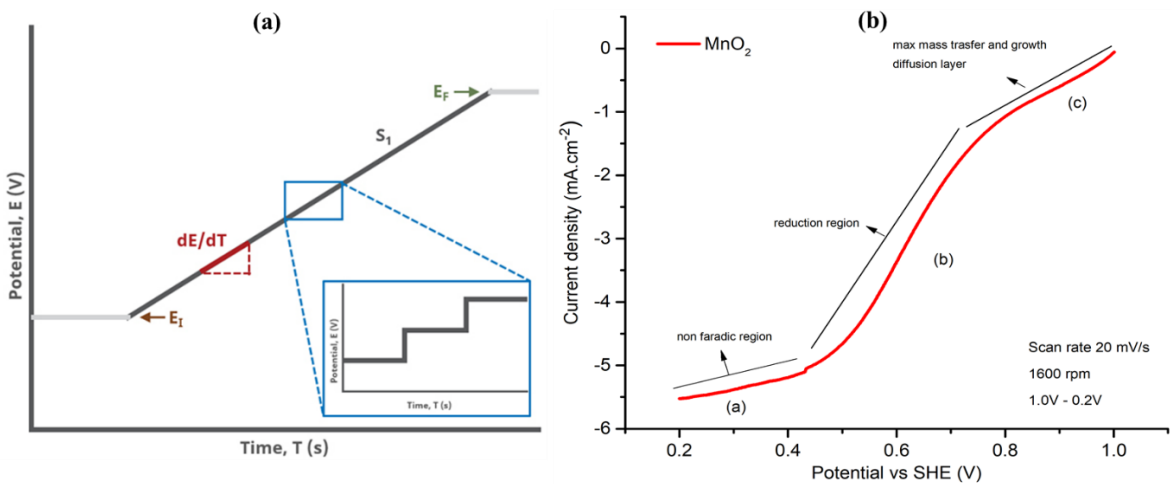

**Figure 8.** *Linear sweep voltammetry (LSV) (a) Excitation signal of the linear scan, (b) reduction voltammogram*

Considering equation (10), with reduction potential $E_o$ and the electrode potential $E$. If the potential of working electrode $E$ starts more positive then reduction potential $E_o$ reactants $A^+$ will not reduce hence a non-faradic current will flow in the system till $E = E_o$ (point (a) in figure 8b). As the working electrode potential in the RDE system reaches the reduction potential ($E_o$) of the $A^+$ reduction process will start and $A^+$ will start reducing to A. The reduction process will create a gradient near the electrode surface due to which will lead to an increase in the mass transfer of electroactive species in electrolyte and there will be an enormous increase in the faradic current as $E$ increases in a negative direction (b). In the meantime, an increase in mass transfer will also build a yield of reduced product $A$ near the working electrode. At $E>E_o$ the mass transfer will reach its maximum and nearly all $A^+$ reactants will have reduced to product $A$ the slope of the I-V plot will begin to tail as $E$ has swept to its final value $E_f$. The resulting current recorded at that point will be known as limiting current ($I_{lim}$) which is the max cathodic current recorded for the examining material. The value of $I_{lim}$ is defined by equations 11 and 12 respectively. It is important to know that for the same reactants $A^+$, $I_{lim}$ increases with an increase in rpm of the working electrode in the RDE system as it increases the convection of electroactive species till all the species have been reduced to $A$.

Like a CV, LSV is a powerful characterization tool for determining the concentration and origin of unknown species. Each reactant species has a unique value of electrode half potential *($E_{1/2}$)* through which we can identify unknown species while the height of $I_{lim}$ gives us information about concentration. LSV technique is used mostly in an irreversible electrochemical reaction where CV is very limited. For example, the bioelectrochemical reduction of $CO_2$ to methane is an irreversible process and a higher limiting current density is required to produce methane, therefore, LSV has a distinct advantage over CV.

### 2.2.3. Tafel Slope analysis

Tafel slope analysis is used in electrochemical systems to determine the rate of reaction which is associated with the mass transfer process. Tafel analysis shows a direct relationship between the applied electrode potential and the reaction rate which simply tells how efficiently a change in potential on the working electrode generates an electric current. In other words, it shows the sensitivity of electrode material (electrocatalyst) in electrolyte towards excitation signal [18]. The Butler-Volmer equation.

$$I = I_o \, exp[(\acute{a}nF/2.3RT) \, \eta] \tag{17}$$

Where $I$ = current density (A/m$^2$), $I_o$ = exchange current density (current density at equilibrium potential between both electrodes), $\acute{a}$ = charge transfer coefficient, $n$ = no of electrons transfer in reaction, $F$ = faraday constant, $R$= universal gas constant (J/mol.k), $T$ = temperature (K), $\eta = (E - E_{eq})$ = activation overpotential, $E_{eq}$ = Potential of electrode at equilibrium.
Rearranging the equation.

$$\eta = (\acute{a}nF/2.3RT).log \, (I/I_o) \tag{18}$$

In equation (17) the term $(\acute{a}nF/2.3RT)$ = constant and is called **Tafel Slope** (mV/decade). The slope mainly depends on the charge transfer coefficient $(\acute{a})$ which is the measure of the energy barrier in the energy coordinate diagram. Therefore a change in slope will change the rate of reaction. A direct relation between $(\eta)$ and $(\acute{a})$ means that the high overpotential $(E - E_{eq})$ slope will be higher which means that the reaction rate will be slow and vice versa [19]. So in fuel cell and metal-air batteries where oxygen reduction reaction (ORR) and oxygen evolution reaction (OER) is of critical importance, we will require electrode material that gives high current densities $(I)$ at lower overpotential $(\eta = E - E_{eq})$ i.e. small Tafel slope.

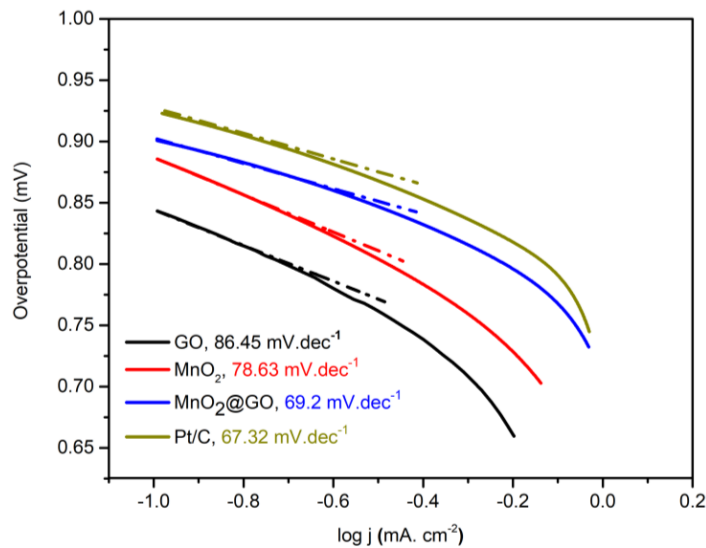

**Figure 9.** *Tafel slopes of cathode materials used in Zn-air batteries and alkaline fuel cell*

The Tafel slopes in figure 9 can be generated from LSV analysis by taking the absolute value of current density *($I_{abs}$)* at 1600 rpm and plotting it on logarithmic x-axis coordinate against electrode potential.

## 2.3. Chronoamperometry

Chronoamperometry is another electrochemical technique used in batteries and fuel cells to study electrode reaction mechanisms. In this technique, a faradic current generated at the working electrode is monitored as a function of time by applying a single or double step excitation (potential) signal. Generally in a chronoamperometric experiment, the potential *E* of the working electrode in RDE setup is first kept above $E_o$ at which no faradic current occurs. The potential is stepped to a value at $E > E_o$ in zero seconds where the oxidation or reduction process occurs [20]. Taking an example of reduction process of solid working electrode material immersed in an electrolyte containing oxidized species (oxidants) of redox couple at a known concentration ($C_o$). At the start of an experiment when $E<E_o$ there is a high concentration of oxidants. Most important to note that the process should be strictly diffusion therefore the electrolyte should be stirred at constant rotational speed during the process. As the potential is stepped up $E > E_o$ the oxidants start reducing immediately. In very little time all the concentration of oxidants near the electrode surface falls to zero thus creating a concentration gradient $\partial C/\partial x$. As a response, the oxidants far away from the electrode surface diffuses to the electrode surface and get reduced. As long as the working electrode remains at this potential the *diffusion layer* that has been created due to the concentration gradient extends farther away from the electrode surface as a result slope of the concentration gradient $\partial C/\partial x$ decreases thus resulting in the decay of generated current (*I*). The same process applies to oxidation in which the gradient generated by oxidized species is in opposite direction. Figure 10 shows the potential stepping and chronoamperometric responses for both oxidation and reduction processes.

Chronoamperometric analysis are based on the standard Cottrell equation which defines the decay of electrode current *(I)* as a function of time [21].

$$I(t) = nFACD^{1/2}/(\pi t)^{1/2} \qquad (19)$$

The Cottrell equation is based on the fact that current measured during redox activity depends on mass transport of reactant through diffusion process. Therefore the current measured during a uniform electrode potential is said to be a diffusion-controlled current.

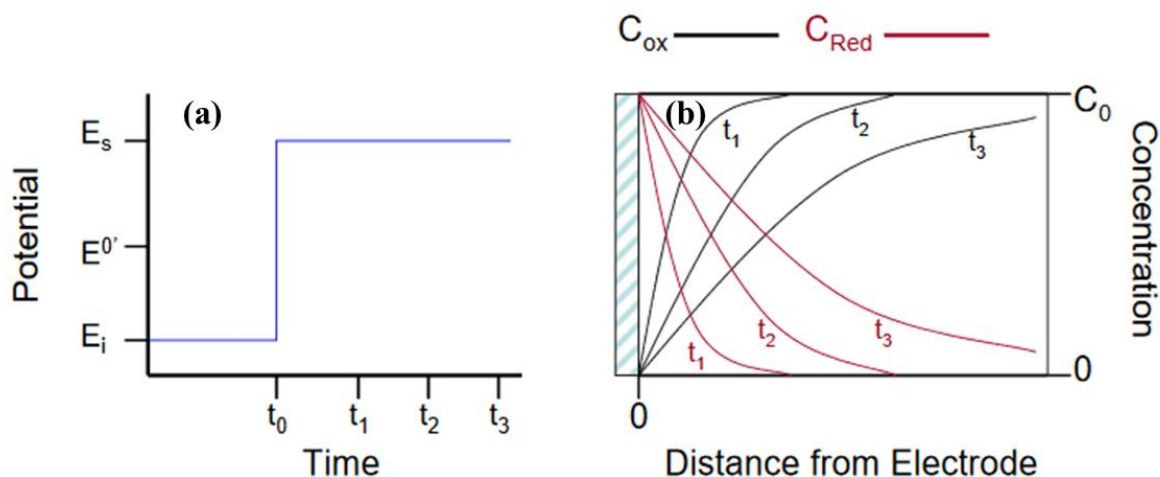

**Figure 10.** *Chronoamperometry (a) Single-step potential, (b) Growth of diffusion layer in redox chronoamperometry* [22]

Chronoamperometry is a very effective method to study different electrode materials by recording their current response w.r.t time at constant electrode overpotential. The properties associated with different electrode materials such as adsorption of oxygen ions over electrode surface in ORR, number of electron transfer ($n$), electrode stability/durability, Diffusion coefficient ($D$), and surface area of working electrode if $C$, $D$ at $n$ are known. Double step (oxidative and reductive) chronoamperometric analysis (figure 11) is used in some cases to determine the reversibility of a chemical reaction by comparing the current responses from two potential steps [23].

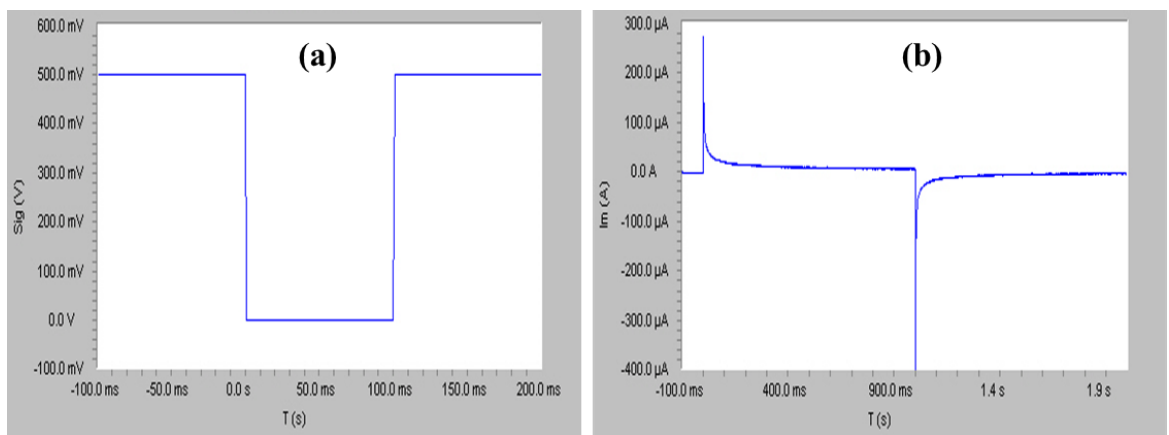

**Figure 11.** *Chronoamperometry (a) double step potential, (b) current response* [23]

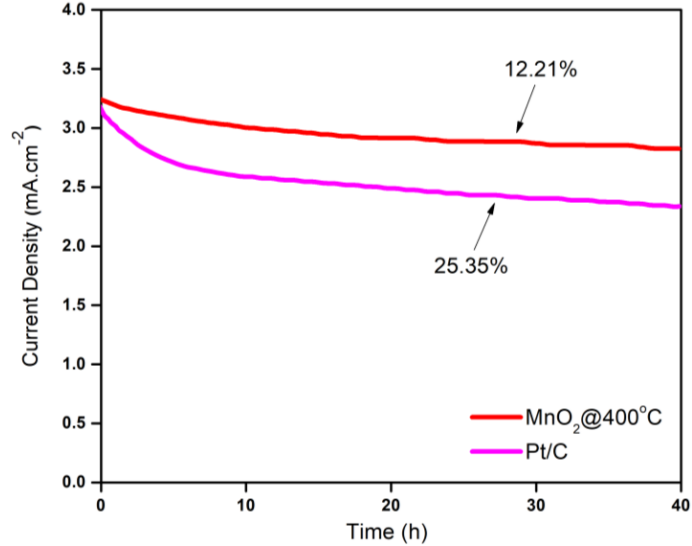

**Figure 12.** *Chronoamperometric response of MnO₂ and commercial Pt/C* [24]

Figure12 compares chronoamperometry analysis of manganese dioxide sintered at 400°C with commercial Pt/C (20%). The analysis was carried out at 0.5V and 400 rpm for 40 hours in 1M KOH electrolyte solution. There was a 25% reduction in current density recorded for Pt/C whereas under the same conditions MnO₂@400°C was more stable in the reduction environment and showed only a 12% reduction. The reduction of MnO₂@400°C was as much half as compare to commercial Pt/C.

### 2.4. Chronopotentiometry

Chronopotentiometry is a current step technique which is applied between working and counter electrode. A constant current is applied at the working electrode of RDE/FC and the potential changes to a value at which the flux of electroactive species are sufficient enough to supply the applied current [25]. After some time the flux of the electroactive species near the electrode surface in the electrolyte cannot sustain the same current due to a change in concentration (redox reaction) and the voltage jumps to a lower value at which other species are reduced. So basically in chronopotentiometry, the current is input and voltage is output. The transition time ($T$) required to change from one voltage value to another is described by the Sands equation.

$$T = ((\pi)^{1/2} n\, F\, D^{1/2} C^*)^{1/2} / (2 i_o)^{1/2} \qquad (20)$$

Where n represents the number of electrons transfer during a reaction, F is Faraday constant, $C^*$ is the concentration of electroactive species, $i_o$ is the current density and D is the diffusion coefficient.

The current-voltage response obtained at different step currents is plotted on x-y coordinates called polarization curve (Figure 13).

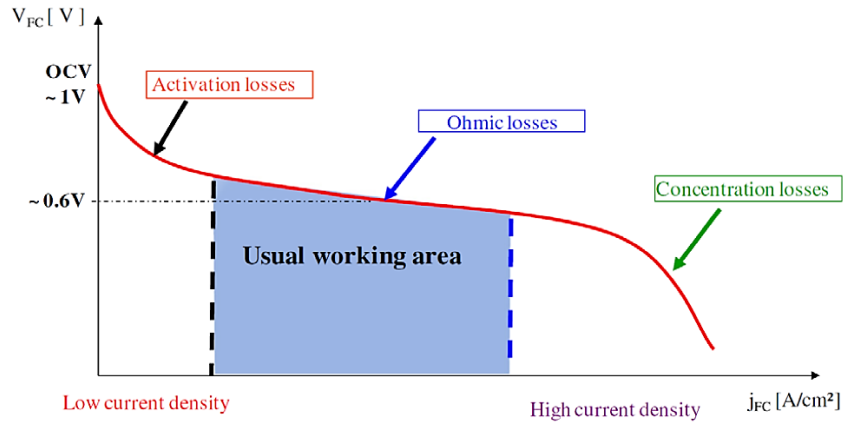

**Figure 13.** *Polarization plot for hydrogen fuel cell system*

The first point at voltage coordinate is called open-circuit voltage (OCV) at which no load is connected to withdraw current from the cell. The three identified regions in figure 13 are related to the voltage drops resulted from three main losses in fuel cell systems. The activation loss at small current densities is caused by sluggish electrode (typically cathode) kinetics of fuel cells. This is the place where highly active catalysts are required on electrode surfaces that can break energy barriers easily thus minimizing the activation losses. The second region where the I-V curve follows a linear relation is called the ohmic losses region and this is the resistance offered by fuel cell membrane electrical circuit connections, current collector, and bipolar plates towards the conduction of protons and electrons. The final region that shows a sharp drop in cell voltage at high current densities is called concentration losses which are due to the mass transport resistance specifically resistance to Oxygen in GDL in FC [26].

Chronopotentiometry analysis is an effective electrochemical characterization method to study the capacity of a fuel cell involving membrane electrode assembly (MEA) to provide accurate voltage response for a given fuel cell current. It provides information about the best operating conditions of FC relating to different parameters such as relative humidity (RH), best operating pressure (P) and cell temperature (T), etc. Chronopotentiometry of FC's is measured in galvanostatic mode (current step) to measure the voltage of the cell.

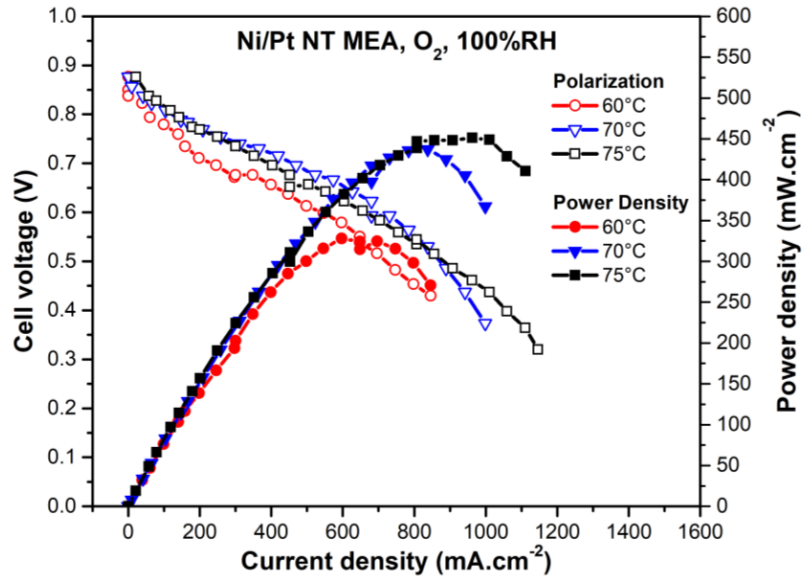

**Figure14.** *PEM fuel cell polarization curve using Ni-Pt as cathode catalyst* [27]

## 2.5. Electrochemical Impedance Spectroscopy (EIS)

Electrochemical Impedance Spectroscopy (EIS) is a powerful electrochemical characterization technique used for characterizing solid-electrolyte interfaces in electrochemical systems e.g. batteries and fuel cells. Initially, EIS was used as a laboratory technique for finding double layer capacity (metal-electrolyte interfaces) but with time its application became wider on an industrial scale such as metal corrosion [28,29], electrodeposition [30,31], and characterizing electrical properties of electrode materials [32–34]. With advancements in the fuel cell, EIS has been widely used for characterizing internal resistance and diagnosing its electrical properties [35].

EIS is a non-destructive technique that uses an AC signal (~10 mV) of very small amplitude from a potentiostat superimposed over the DC from the characterizing source at various frequency ranges (1 MHz – 100 kHz). The response obtained from characterizing the source is received by the EIS system from which the graph of a signal can be obtained. Typically Nyquist and Bode's plots are used to analyzing the response from the system. But most commonly Nyquist plots are used which are composed of a relationship between imaginary impedance resistance (Y-axis) and real impedance resistance (X-axis) as shown in figure 15 [36].

As explained in chronopotentiometry, fuel cell systems suffer from three main types of voltage losses: Activation losses (charge transfer activation), Ohmic losses (Proton and electron transport), and concentration losses (mass transfer). EIS can be used to quantify these losses and distinguished between them. EIS can be operated both in voltage control (potentiostatic) and current control (galvanostatic) mode. In potentiostatic mode, AC signal (~10 mV) is applied to disturb the electrochemical system, and the response obtained from the characterizing source is plotted on Nyquist plot and vice versa. For different interfaces, EIS data can be modeled through software's (Biologic[TM] EC lab) to fit it in various equivalent circuits e.g. Figure 15 shows the EIS spectrum of PEM fuel cell membrane electrode assembly (MEA) composed of resistor (R), capacitor (C) and inductor (I) connected in series and parallel combination. The equivalent circuit shown in figure 15 was used to fit the Nyquist plot. The resistance values that intercept the real axis represents ohmic resistance and is mainly composed of membrane resistance ($R_m$) and other two resistances i.e. Charge transfer resistance at the anode ($R_a$) and cathode ($R_c$) whereas $C_a$ and $C_c$ are capacitance accompanying $R_m$ and $R_C$ respectively.

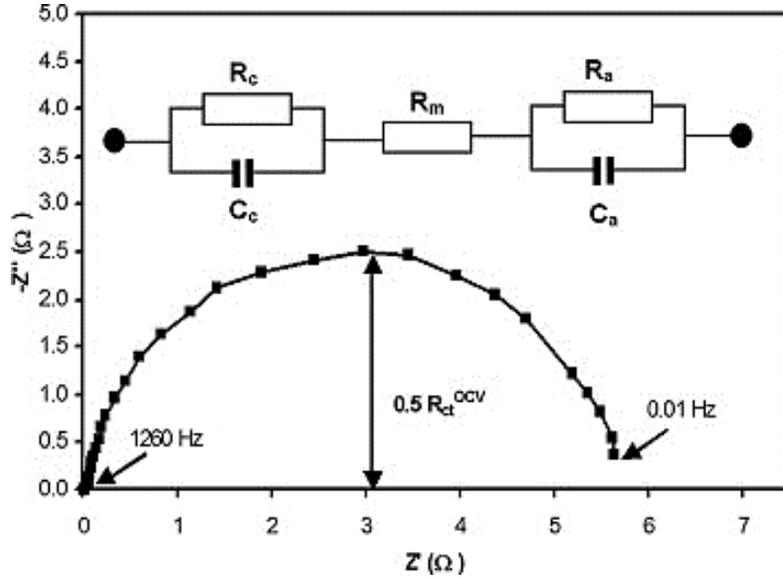

**Figure 15.** *EIS Nyquist plot of PEM fuel cell at OCV, 80°C with Nafion 112 membrane, $H_2$ and $O_2$ supply and 100%RH* [36]

In hydrogen fuel cells the kinetics of hydrogen oxidation reaction (HOR) occurring at the anode is much faster than oxygen reduction reaction (ORR) at cathode $R_c$, ($R_a \ll R_c$) therefore $R_a$ can be ignored and $R_c$ can be treated as $R_{\text{charge transfer}}$, ($R_c = R_{ct}$) and resistance for charge transfer can be found by differentiating famous Butler-Volmer equation.

Butler-Volmer equation for fuel cell OCV overpotential.

$$\eta = (RT / n F\, i_{O2})*I \qquad (20)$$

where η is overpotential at OCV, R is the universal gas constant, n is the electron transfer number, F faraday constant $i_{O2}$ apparent exchange current and I is the current density. So based on the above discussion as Impedance is performed by very small AC ripple excitation.
Therefore.

$$R_{ct} = \partial\eta/\partial I *(RT / n F\, i_{O2})*I \qquad (21)$$

According to the figure as the maximum Y, value is half of charge transfer resistance and $R_a$ is ignored therefore

$$R_c = 0.5R_{ct} = 0.5*(2.5) = 1.25\,\Omega$$

And

$$R_m = R - R_{ct} = 5.7\,\Omega - 1.25\,\Omega = 4.4\,\Omega$$

## 3. Conclusion

To conclude, the report has presented a thorough study on the common battery testing method used in fuel cell research. All corrosion tests can be performed through RDE which is a tri electrode system connected to a potentiostat. Voltammetric methods are used to determine the kinetics of redox reaction, number of electron transfer, reaction reversibility, current density, cell potential, and adsorption for a metal electrode. Chronoamperometry analysis is used to determine the durability of the electrode through degrading current by subjecting it to the constant potential for hours. Chronopotentiometry analysis determines the operating capacity and optimum operating conditions for working for electrode assembly by applying constant current and recording the change in applied potential. EIS techniques characterize the electrical conductivity of membrane and electrode assembly by finding the membrane and cathode resistance by superimposing small AC pulse excitation signal from the potentiostat on the DC from the source.

**References.**